\documentclass{article}

\usepackage{graphicx}
\usepackage{verbatim}
\usepackage{authblk}

\title{A graph neural network-based approach to XANES data analysis}
\author[1,2]{Fei Zhan}
\author[1]{Lirong Zheng}
\author[1]{Haodong Yao}
\author[1]{Zhi Geng}
\author[1]{Can Yu}
\author[1]{Xue Han}
\author[1]{Xueqi Song}
\author[1]{Shuguang Chen}
\author[1]{Haifeng Zhao\thanks{zhaohf@ihep.ac.cn}}
\affil[1]{Institute of high energy physics,Chinese academy of sciences}
\affil[2]{Dongguan Institute of Neutron Science, Dongguan 523808, China}

\begin{document}

\maketitle
	
\begin{abstract}

\quad X-ray absorption spectroscopy (XAS) is an indispensable tool to characterize the atomic-scale local three-dimensional structure(3D structure) of the system, in which XANES is the most important part.
The quantitative analysis of 3D structure from XANES however, requires users to have a deep understanding and accurate judgment of cluster and summarize the most important structural variables, which is often difficult to achieve. 
In this work, we construct a graph neural network model to simulate XANES, 
it takes the 3D structure of system as input while the inherent relation between the fine structure of spectrum and local geometry can be considered during the model construction; It turns to be faster than the traditional XANES fitting method when we combine the simulation model and XANES optimization algorithm together to fit the 3D structure of the given system. Our method does not require the summarization of the structural variables for the sample, so it can be used in the structure characterization for any solid materials, it has positive significance for the study of structure-function relationship in the fields of energy and catalysis. In addition, the method is expected to be the key part of online 3D structure analysis framework for the XAS related beamlines of high energy photon source(HEPS) under construction.

\end{abstract}	
	
	\par
	Keywords:X-ray Absorption Spectroscopy, XANES, Machine Learning, Graph neural network, Optimization Algorithm
	
\section{Introduction} 
	
\par
X-ray absorption spectroscopy (XAS) is one of the most powerful tools to detect the local structure around the absorber in kinds of materials. Given the need to quickly characterize the spectra, there is a growing interest in developing efficient methods for XAS analysis. 
Machine learning has emerged as a powerful tool for fast data analysis due to its ability to automatically learn patterns and relationships in large datasets. Unlike traditional analytical methods that require domain expertise, machine learning can identify relevant features and make predictions based on data-driven insights. Futhermore, after the completion training of machine learning model, the time required for individual model predictions is trivial and noticeably less than that of conventional data analysis. The automation capabilities,low time consumption and low user expertise requirements with machine learning make it an ideal choice for fast data analysis in XAS.
In 2017, Timoshenko et al. first introduced machine learning into XAS analysis\cite{timoshenko2017JPCL}, they retrieved the coordination number, radial distribution function\cite{timoshenko2018PRL} and coordination bond length\cite{timoshenko2020bond} of absorber in the metal clusters with fully connected network(FCN). 
Frenkel et al. went further to the bimetallic sites nanoclusters and combined two related FCNs to analyze the coordination number of both sites together\cite{li2021frenkel}.
Zheng et al. applied another popular model the random forest to analyze the coordination mode identifier of the system\cite{zheng2020patterns}. Presently, the outputs of machine learning models are constantly enriched, from the initial coordination number and bond length to the radial function and coordination motifs.
Among various structural information, 3D structure is of most importance, which plays a key role in revealing the microscopic mechanism of material strain, chemical reaction mechanism and in-depth study of the relationship between structure and function of the system. X-ray absorption near edge structure (XANES) is sensitive to the local geometry around the absorber then is well used to probe the 3D structure of given system. Due to the complexity and time-consuming nature of XANES simulation, relatively few studies have been conducted in quantitative 3D structure analysis of XANES. 
The extended X-ray absorption fine structre(EXAFS), however, dominated by the radial structure (two dimensional) around the absorber, is much easier to treat due to the well known single scattering equation\cite{rehr2000MS}, so its fitting accounts for the majority of quantitative analysis work in XAS. 
To get more detailed atomic geometry in the system,  employing machine learning models to analysis  the 3D structure of a given system is encouraging, it is necessary to design a 3D structure output coding scheme. To our best knowledge, there has been no paper published model that takes the 3D structure of the system as the direct output.
\par
As noted, XANES fitting is the pivotal technique to derive the 3D structure from XAS. In the fitting, there is always one module to calculate the spectrum, which we call "core" for short,  while the fitting of the structural parameters is usually controlled by optimization algorithms.The "core" module used to produce the spectrum can be either based on multiple scattering(MS) treatment\cite{natoli1986} or by the machine learning model.
The former has a long history and is taken as the classic choice for interpretation of XAS.
MXAN\cite{MXAN}, a program developed by the Benfatt’s group, integrates CONTINUUM for MS calculations with the MINUIT optimization package. It is the most widely used software for local 3D structure analysis of metal protein sites from XANES,such as PML protein\cite{MXANapp}. The XANES fitting method developed by Rehr’s group employs the FEFF code for MS calculations, utilizing a Bayesian optimization algorithm\cite{rehr2005bayes}, and it was applied to PtPOP complex\cite{veen}. We introduced a deterministic global optimization algorithm in the fitting loop while the core is FEFF or FDMNES\cite{zhan2017alternative}. Our method offers an additional fitting of difference spectra for time-resolved studies, such as photocatalytic $CO_2$ reduction system of Ni complexes\cite{hu2020}. 
With the advancement of artificial intelligence emerging in big data processing frontier, several packages have been developed to that adopt machine learning model for XANES fitting in recent years. By utilizing the machine learning model for XANES calculations, it can significantly reduce the time consumption associated with theoretical simulation. These packages includes Fitit\cite{FITIT}, based on interpolation model and futher Pyfitit\cite{pyfitit}, based on regression models such as random forest etc. Algorithms like random forests maintain high computational efficiency without relying on experience and trial to gradually build interpolation polynomials, which can be user-friendly. Smolentsev et al. used their software FITIT to study the system of CoII(aPPy) catalytic decomposition of water to produce hydrogen, and analyzed the four-nitrogen atom coordination intermediate \cite{smolentsev2018}.
Obviously, the three dimensional geometry of complex and metalloprotein system is the focus of XANES fitting application. 
\par
Given the widespread application of XAS, in the quantitative data analysis of XANES, researchers may encounter diverse systems. An approach that involves the direct fitting of 3D coordinates demonstrates broad applicability.
Therefore,the development of a machine learning model capable of efficiently and precisely predicting XANES by inputing atomic coordinates directly is imperative, as this can significantly accelerate the fitting process. 
Previous machine learning algorithms are not convenient for inputting 3D coordinates, it is a feasible technical route to improve the ability of machine learning model by inputting additional 3D coordinates into the model then improve the consistency between the model predicted spectra and theoretical calculated ones. Recently Kotobi et al. applied GNN to predict C K-edge XAS from organic molecules 3D structures\cite{GNNJACS}.
3D graph neural network(3D GNN) is a good choice, within this framework, Graphs serves as potent instruments for modeling high-degree systems, has been used in molecules, proteins and materials\cite{xie2018crystalgnn}, quantum systems\cite{fu2022lattice}. Among various types of machine learning models, 3D GNN\cite{schnet2018}, specially designed to deal with 3D coordinates input of the system, if needed, with atoms as nodes and atom pairs as edges, is considered to be the natural choice for 3D structural analysis. The differences between different 3D GNN models such as SchNet\cite{schnet2018}, SphereNet\cite{spherenet2021},DimeNet++\cite{dimenet2020} lie in the completeness of the definition of geometric features, the computational efficiency of geometric features and the definition of characteristic functions. Here, we develope an alternative 3DGNN model tailored for XANES prediction, which takes into account the varying significance of various geometric features and futhermore the essential role of the local environment around absorbing atoms in XANES prediction. It is proved to be more efficient in XANES simulation than the current GNN models.
\par
The manuscript is structured as follows. A novel XANES fitting method that utilizes a customized 3D GNN model, XAS3D, for XANES calculation will be introduced in section 2. The impact of diverse geometric attributes and graph definitions on the Graph Neural Network (GNN) model is scrutinized, and the model is tailored accordingly. Section 3 focuses on two practical applications of the method. The $Fe_3O_4$ system exhibiting two distinct coordination environments, tetrahedral and octahedral, makes it an appropriate choice for evaluating the performance of the XAS3D model on complex systems, which will be checked in detail in section 3.1, the application of our XANES fitting method in a Mn-doped $Co_3O_4$ system will be given in section 3.2. The further development of this method is prospected in the section 4.

\section{Method} 

\begin{figure}
	\centering
	\includegraphics[width=0.7\linewidth]{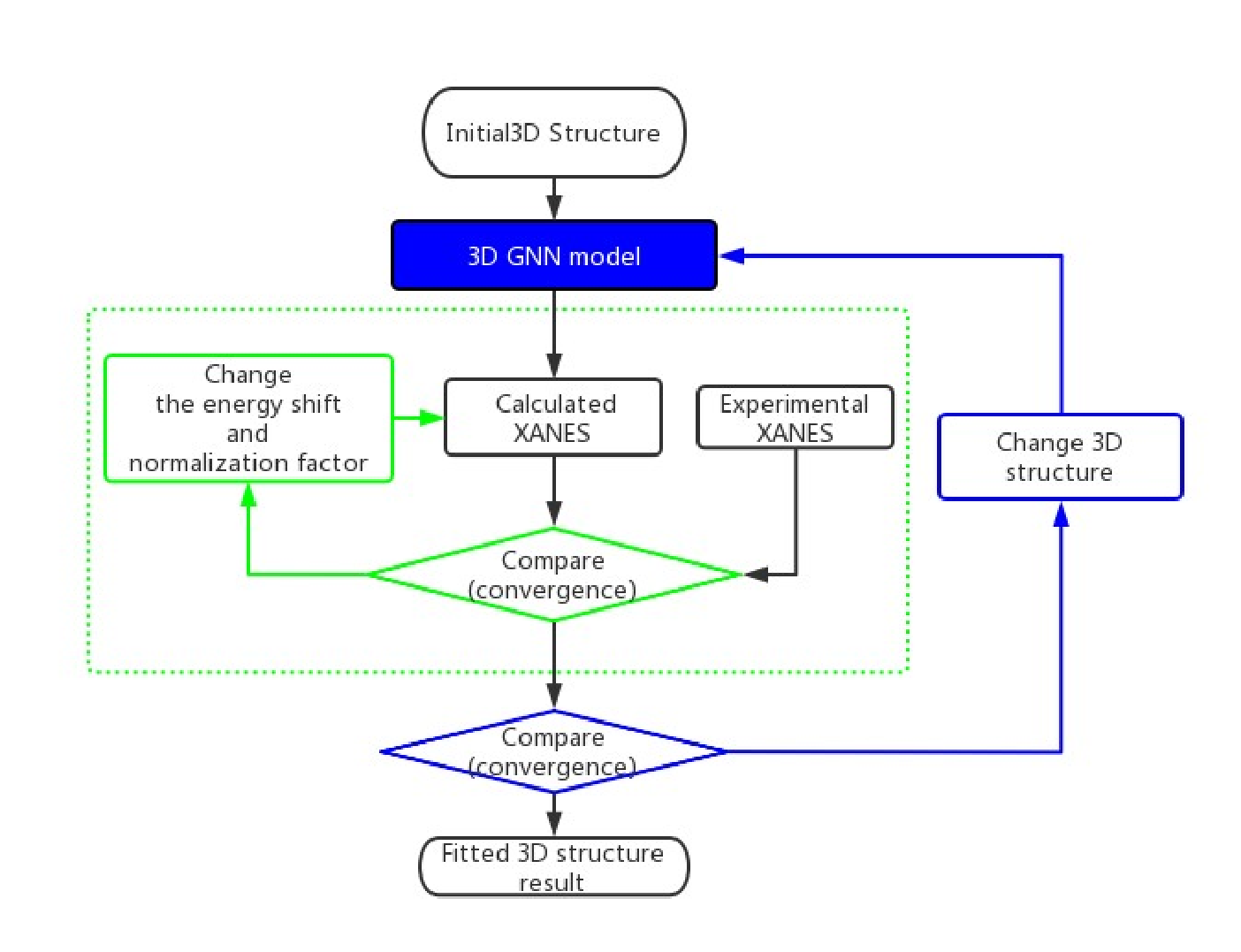}
	\caption[short]{The flowchart of XANES fit combined GNN model and optimization algorithm.}
	\label{fig:flowxanesfit}
\end{figure}

\begin{figure}
	\centering
	\includegraphics[width=0.7\linewidth]{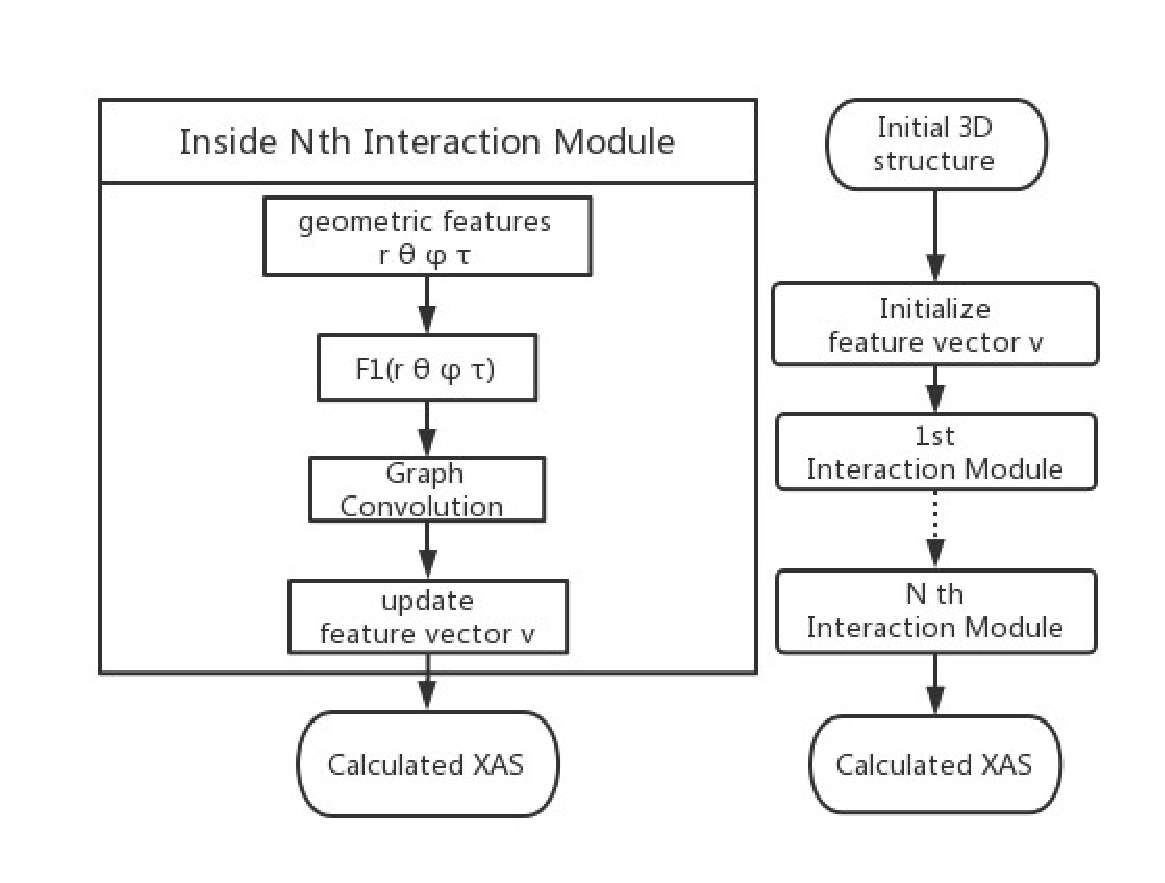}
	\caption[short]{The flowchart of XAS3D GNN model.}
	\label{fig:flowgnn}
\end{figure}

\par

In this section, we show an innovative method suitable for solid materials and nanoclusters. These systems lack effective rigid constraints that reduce the degrees of freedom for fitting, such as peptide chains in proteins and benzene, pyridine rings in complexes, thus requiring the handling of more 3D coordinates. The flowchart of the method is shown in the Fig.\ref{fig:flowxanesfit}. 
A 3D GNN model XAS3D is constructed and trained, then taken as a calculator to produce a XANES quickly by inputing the 3D structure of given cluster, the simulated spectrum will be aligned and normalized to meet the best of agreement with the experimental measure, the structure of cluster will be modified and reinput into the XAS3D model to generate the new spectrum if the coincidence between simulated spectrum and experimental one are not desirable. The fitting will not stop until it meets the goodness of agreement or the loop-jump-out settings in the optimization algorithm. Note for the mental complex, the 3D cartesian coordinates of atoms in the cluster are directly used as input for XAS3D, while for solid materials, the lattice parameters and fractional coordinates will be first converted into cartesian coordinates of a cluster centered on the absorber atoms, then be used as input for XAS3D. 
For general materials, including but not limited to mental complex, protein, the 3D cartesian coordinates are directly used as input for XAS3D. Moreover, for crystals, the lattice parameters and factional coordinates coordinates are first converted into cartesian coordinates of a cluster centered on the absorber atoms, which are then used as input for XAS3D. The proposed method employs a custom-designed 3D GNN XAS3D to calculate the XANES spectra of a given structure. The XAS3D model predicted XANES are compared with the experimental spectra. The fitting process is terminated when the difference between the two meets the stopping criteria of the optimization algorithm; otherwise, the structure is changed for another round of XANES fitting step. During the comparison of predicted XANES and experimental one, the normalization factor and energy shift are continuously adjusted, making the comparison process a parameter fitting as well.
\par
The 3D GNN model XAS3D is the key of XANES fit framework and is constructed in the following way.
We start a 3D graph for a cluster with n=$N_{atom}$, defined as G(V,A,P) where V =[v1;v2;...; vn] is the atom-dependent vector set and will be taken as the node (atom) feature in the graph, the dimensions of all atomic vector vi are same, denoted as $N_{hidden}$. 
The atomic pairs in a graph are represented by an adjacency matrix A. There is an edge $e_{ij}$ in graph G when A[i][j] = 1, $e_{ij}$ is the pair of atoms included in the model. The dimension of the adjacency matrix A is $N_{atom}*N_{atom}$ which is a sparse matrix of dimension $2*N_{edge}$ in the program. The position matrix P consisting of the 3D coordinates of the atoms which is the input of XAS3D model.
In this work, a 3D GNN model XAS3D is developed, which takes 3D coordinates as input and XANES spectrum as output. The main part of XAS3D model is composed of one embedding layer and several interaction layers, and the number of interaction layers is the hyperparameter, as shown in the Fig.\ref{fig:flowgnn}.  The Embedding layer encodes the input atomic species into initial node feature vector v. The following interaction layers updates the node feature vector v based on neighboring node features and geometric features $r,\theta,\phi$ through weighted graph convolution. The XAS3D model takes the interaction layer as the fundamental unit, which is detailed in the magnified left inset of Fig.\ref{fig:flowgnn}. The interaction layer first calculates the distance r, angle $\theta$, dihedral angle $\phi$ and other geometric features corresponding to the atom pairs (e.g. edges in graph) through the input atom coordinates, next calculates the feature function values based on the geometric features, then updates the feature vector of the atom via graph convolution. Finally sum-pooling performs on all atomic node features to obtain the XANES as the predictions of whole graph. 
\par
Subsequently, the geometric features to be included in the graph convolution should be considered carefully, they play a role in the precision of XANES simulation. Generally, the bond lengths r,angles $\theta$ and dihedral angle$\phi$ between absorber and the surrounding atoms dominate the fine structures in the spectrum, so they are the most efficient geometric features to be chosen for the model utilizing the calculation scheme from \cite{comenet2022}, the feature function $TBF(r,\theta,\phi)$ is selected to enhance the model’s efficiency in ComENet,SphereNet and GemNe\cite{comenet2022,spherenet2021,gemnet2021}. For the system where bond lengths r and torsional angles $\tau$ are crucial to the oscillator behaviour near the edge in the spectrum, feature functions $SBF(r,\tau)$ can be employed. In previous XAS3D GNN model, the typical approach involves selecting atom j in the vicinity of atom i as its neighboring atom, based on a predefined uniform distance cutoff. 
But for XANES which primarily investigates the local environment surrounding the absorber atom, the significance of the edges in graph associated with the absorber becomes particularly pronounced, the absorber atom and its neighboring atoms will be extracted to construct the graph in XAS3Dabs model. This graph definition exclusively focused on the local structure surrounding the absorber atom.

\section{Application and discussion} 
\par
In this section, we first validated the effectiveness of the XAS3D model in $Fe_3O_4$. Subsequently, in the Mn doped $Co_3O_4$, we focused on describing the XAS3D model based XANES fitting results. The model construction and performance description for XANES simulation of Mn doped $Co_3O_4$ are detailed in the support information to avoid repetition.
\subsection{Application to $Fe_3O_4$}

\par
\begin{table} 
	\caption{Hyperparameters used in the XAS3D model and their values.}
	
	\begin{tabular}{lccc}
		\hline
		Model Hyperparameters                                &Abbreviation  & Values in grid search & Value Range in sampling  \\
		\hline
		Number of layers                                     &n\_l          & 3           &1-3 \\
		Hidden embedding size                                &hidden        & 64,128,256  &18-256  \\
		Middle embedding size                                &middle        & 64,128,256  &18-256 \\
		Number of layers for output blocks                   &n\_out\_l     & 2,3,4       &1-12  \\
		Distance embedding dim                               &n\_r          & 3,6,12      &1-12  \\
		Angle embedding dim                                  &n\_sph        & 3,6,12      &2-7  \\
		\hline
	\end{tabular}
	\label{tbl:hyper}
\end{table}

\begin{table} 
	\caption{MAE statistics of $Fe_3O_4$ XANES prediction with XAS3D and XAS3Dabs models adopting latin hypercube sampled hyperparameters.}
	
	\begin{tabular}{lccccc}
		\hline
		Model                       &Mean of MAE &Median of MAE &Minimum of MAE  &Maximum of MAE   \\                          
		\hline
		XAS3Dabs             &0.0096 &0.0095  &0.0084 &0.0127 \\
		XAS3D                       &0.0186 &0.0140  &0.0098 &0.0419\\      
		
		\hline
	\end{tabular}
	\label{tbl:femaehyper}
\end{table}

\begin{figure}
	\centering
	\includegraphics[width=0.7\linewidth]{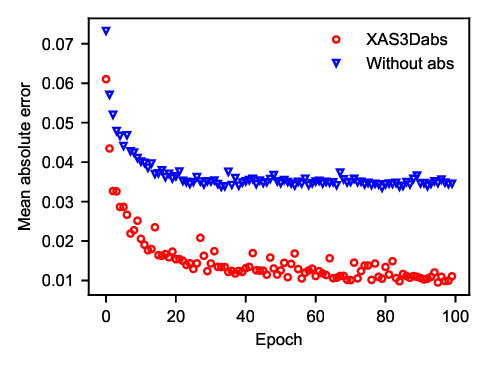}
	\caption[short]{Training curves of model XAS3Dabs and Withoutabs constructed for $Fe_3O_4$ XANES prediction. XAS3Dabs stands for the GNN model including only the absorber related edges, while Withoutabs is the model excluding edges related to absorber atoms.}
	\label{fig:feabs}
\end{figure}

\begin{figure}
	\centering
	\includegraphics[width=0.7\linewidth]{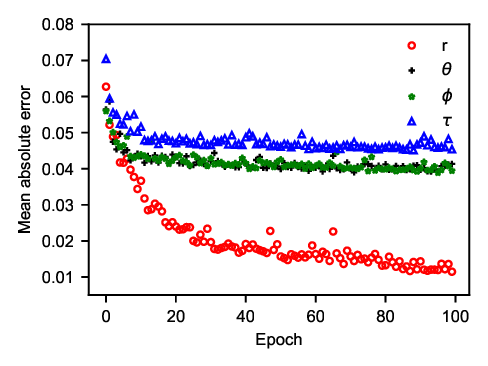}
	\caption[short]{Training curve of 3D GNN models using different geometric features in $Fe_3O_4$.}
	\label{fig:ferthetaphi}
\end{figure}

\begin{figure}
	\centering
	\includegraphics[width=0.7\linewidth]{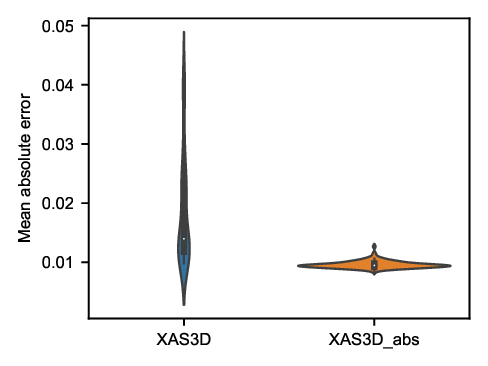}
	\caption[short]{Boxplots of performance metric (MAE) of XAS3D and XAS3Dabs GNN models using different hyperparameters in XANES prediction of $Fe_3O_4$.}
	\label{fig:feboxhyper}
\end{figure}

\begin{figure}
	\centering
	\includegraphics[width=0.7\linewidth]{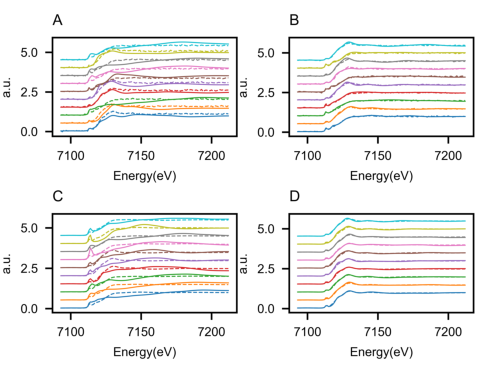}
	\caption[short]{XANES simulated with multiple scattering framework(solid lines) and with machine learning model(dashed lines) on testing dataset of $Fe_3O_4$. A:The 10 worst predicted XANES by multilayer perceptrons. B:The 10 best predicted XANES by multilayer perceptrons. C:The 10 worst predicted XANES by random forest. D:The 10 best predicted XANES by random forest.}
	\label{fig:fevalidml}
\end{figure}

\begin{figure}
	\centering
	\includegraphics[width=0.7\linewidth]{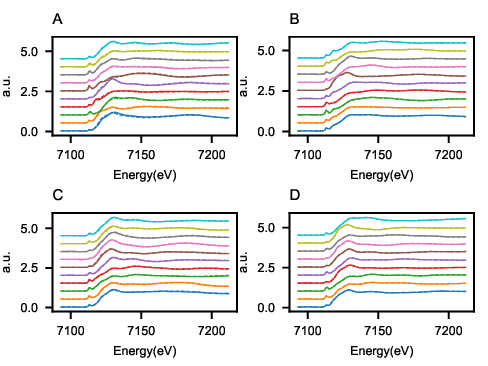}
	\caption[short]{XANES simulated with multiple scattering framework(solid lines) and with machine learning model(dashed lines) on testing dataset of $Fe_3O_4$. A:The 10 worst predicted XANES by XAS3D GNN model. B:The 10 best predicted XANES by XAS3D GNN model. C:The 10 worst predicted XANES by XAS3Dabs GNN model. D:The 10 best predicted XANES by XAS3Dabs GNN model.}
	\label{fig:fevalid}
\end{figure}

\begin{figure}
	\centering
	\includegraphics[width=0.7\linewidth]{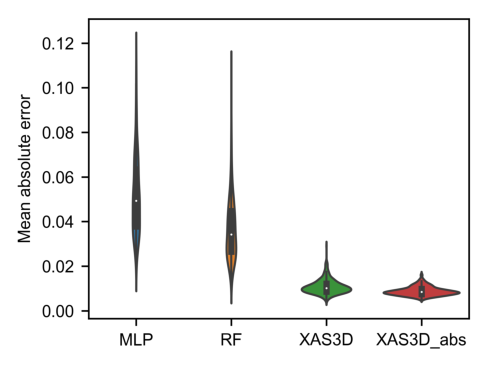}
	\caption[short]{Boxplots of model performance metric (MAE) on testing datasets for $Fe_3O_4$ XANES prediction. MLP:Multilayer perceptron. RF:random forest. XAS3D and XAS3Dabs GNN model.}
	\label{fig:feboxmae}
\end{figure}

\begin{table}
	\caption{Statistical features of performance metrics MAE in the testing dataset for XANES prediction of $Fe_3O_4$ system.}	
	\centering
	\begin{tabular}{lcccc}
		Machine learning model & Mean of MAE & Median of MAE & Minimum of MAE &Maximum of MAE \\
		Multiple perceptron  & 0.0519 & 0.0493    &0.0191 &0.1145 \\
		Random Forest        & 0.0368 & 0.0342    &0.0116 &0.1081\\
		XAS3D                & 0.0108 &0.0102     &0.0045 &0.0292  \\
		XAS3DABS             & 0.0088 &0.0085     &0.0051 &0.0164  \\
	\end{tabular}
	\label{tbl:femae}
\end{table}

\par
Magnetite  $Fe_3O_4$ is one kind of black crystals with magnetism spread of the world, it has distinctive attributes, such as superparamagnetic behavior, minimal toxicity, facile isolation through external magnetic manipulation, $Fe_3O_4$ material manifest substantial promise across a spectrum of applications encompassing the catalysis field \cite{fe3o4_cat1,fe3o4_cat2}, environmental remediation\cite{fe3o4_env}, and bioengineering applications\cite{fe3o4_env}. 
$Fe_3O_4$ system exhibits two distinct coordination environments around the iron: tetrahedral and octahedral. This unique feature makes it an excellent research target for evaluating the capability of a given model to handle complex systems with multiple coordination environments and multiple absorption sites. 
\par
We applied our method to $Fe_3O_4$, where the Fe atom is located in both tetrahedral and octahedral coordination environments.
The $Fe_3O_4$ conventional cell consists of four primitive unit cells and contains 56 atoms of which 24 are Fe atoms as shown in Fig.S4, resulting in a total of 168 coordinates that need to be fitted. 
All the atoms in the conventional cell allows to vary with the range of [-0.3\AA,0.3\AA] in the equilibrium site, and latin hypercube sampling\cite{ihs} was adopted here for its efficient way to coverage parameter space with fewer samples. A total of 200 structures were sampled,leading to 4800  iron-dependent sites, namely clusters centered around the iron absorbers. FDMNES, the most popular XAS simulation package, was adopted here to generate XANES for these iron absorbers in the cluster using its multiple scattering theory calculation mode. Real Hedin-Lundqvist exchange-correlation potential is used in the multiple scattering calculation, and a cluster with 5\AA radius around the absorber is sufficient for the self-consistency in the charge distribution/Dyson equation/Green function iteration loop\cite{joly2009self}.  90$\%$ of the dataset was taken as the training set while the rest was used as the testing set.
Upon commencing the construction of the model of the $Fe_3O_4$ system, we initiated by assessing the significance of geometric features in selecting the feature function for the 3D GNN models.
The training history of models incorporating different geometric features r,$\theta$,$\phi$,$\tau$,is shown in the Fig.\ref{fig:ferthetaphi}. It can be seen that r is the most effective geometric features using in graph convolution layers as edge weights, with the lowest MAE for the model utilizing r as the geometric feature. Additionally, $\theta$ and dihedral angle $\phi$ are more important than torsion angle $\tau$ for $Fe_3O_4$ XANES prediction . 
One possible reason is that $Fe_3O_4$ dataset does not involve torsion angle related issues, such as isomer discrimination\cite{comenet2022}. 
Therefore, based on the test results, the feature function $TBF(r,\theta,\phi)$ is selected to enhance the XAS3D and XAS3Dabs model's training efficiency.
Subsequently, the effect of graph definition on the 3D GNN model was investigated. 
For comparison, we created another graph by removing the atomic pairs (represented as edges in the graph) associated with the absorber atom, and the model utilizing this graph definition was termed Withoutabs. This approach aimed to evaluate the significance of the absorber atom's local structure by excluding it from the graph. To visualize the training progress and compare the performance of models trained on these two graph definitions, we plotted their respective training histories. As illustrated in Fig.\ref{fig:feabs} , a notable difference in model performance can be observed. Specifically, the model trained on the graph definition centered around the absorber atom (XAS3Dabs) demonstrated superior performance compared to the model trained on the graph without the absorber atom related edges(Withoutabs). This finding underscores the crucial importance of the local structure surrounding the absorber atom in determining the overall performance of the XAS3D GNN model. This observation aligns with the inherent physical principles governing X-ray absorption spectroscopy, which is primarily sensitive to the local atomic environment of the absorber atom. The initial construction of the XAS3D and XAS3Dabs models has been completed. 
For the two models, Pytorch\cite{pytorch} and PyTorch Geometric\cite{pyg} is used to implement all related methods. The code references relevant parts of DIG framework\cite{DIG}, however, it maintains flexibility by not relying on DIG.
Adam optimizer is used in the train of 3D GNN for a total of 200 epochs. We set the learning rate at 5E-4 and the learning rate decay factor at 0.5, with a batch size of 32.

\par
To avoid the improper selection of hyperparameters affecting the model performance, the hyperparameters of the model were optimized after model construction. The hyperparameters for both XAS3D and XAS3Dabs models and their respective ranges are listed in Tab.\ref{tbl:hyper}. The Latin Hypercube sampling approach was employed, generating 120 samples of hyperparameters for each model. The Mean Absolute Error (MAE) distributions resulting from these samples are presented in Fig.\ref{fig:feboxhyper}, while related statistical data are provided in Tab.\ref{tbl:femaehyper}. It can be observed that under optimal hyperparameter settings, both models achieve comparable performance, with minimum MAE values of 0.0098 and 0.0084 for XAS3D and XAS3Dabs, respectively. The narrower boxplot of XAS3Dabs indicates superior performance stability in response to hyperparameter fluctuations. This is attributable to its architecture, which consists of artificially selected and absorber-related geometric features that minimize superfluous information and decrease model complexity.

\par
The performance of the GNN model in XANES prediction of the $Fe_3O_4$ was studied after hyperparameter optimization. MLP and RF were selected as comparisons, and their hyperparameters were also optimized as shown in the support information.
The performance of the XAS3D and XAS3Dabs models on the testing dataset is shown in the Fig.\ref{fig:fevalid}. 
It is evident that the GNN models can accurately reconstruct even the 10 spectra with the poorest prediction performance. In contrast, the MLP and RF models struggle to capture the general trend of spectral changes in similar cases, as seen in Fig.\ref{fig:fevalidml}. 
To further illustrate the performance of the different models, Fig.\ref{fig:feboxmae} presents a distribution violin map, while Tab.\ref{tbl:femae} provides statistical data on the mean absolute error (MAE) on the testing datasets.
The XAS3Dabs model significantly outperforms both MLP and RF in terms of MAE.
The XAS3Dabs mean of MAE is reduced by 83.0$\%$ and 76.1$\%$ compared to MLP and RF, respectively, and similarly the median of MAE is reduced by 82.8$\%$ and 75.1$\%$, respectively. These findings indicate that MLP and RF perform poorly in XANES prediction for this system compared to the XAS3Dabs model.
The performance of the two 3D GNN models is quite similar in this system, while XAS3Dabs significantly reduces computational resource demands, especially for GPU memory, by incorporating fewer edges in its graph definition compared to XAS3D. XAS3Dabs slightly outperforms MAE average and MAE median, improving them by 18.5$\%$ and 16.7$\%$, respectively, and exhibiting a significant advantage in MAE maximum, improving it by 43.8$\%$. 
Furthermore, as depicted in Fig.\ref{fig:feboxmae}, the MAE distribution of XAS3Dabs is relatively compact, exhibiting a smaller interquartile range and a narrower boxplot, which suggests that its MAE is more consistent across various sample data points.
These results suggest that 3D GNN models, particularly XAS3Dabs, are well-suited for accurate XANES prediction for complexes.
\par
The XANES calculations, which are based on the known structure of $Fe_3O_4$ as the standard sample, are presented in in support information Fig.S2. This section focuses on verifying the feasibility of the XAS3D and XAS3Dabs graph neural network model.

\subsection{Application to Mn doped $Co_3O_4$}
\par
\begin{figure}
	\centering
	\includegraphics[width=0.7\linewidth]{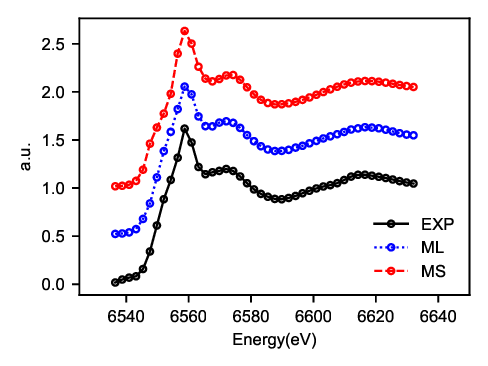}
	\caption[short]{Experimental XANES(EXP), GNN predicted XANES fitting result(ML) and Multiple scattering calculated XANES based on fitted structure(MS) of Mn doped $Co_3O_4$. Reproduced and adapted with permission from ref25. Copyright 2020 American Chemical Society.}
	\label{fig:mnxanes}
\end{figure}

Mn doped $Co_3O_4$ via $Mn^{3+}$ substituting for $Co^{3+}$, unsaturated $Mn^{3+}$ with J-T distortion acts as an active site\cite{JT} and rapidly reacts with groups adsorbed in its vicinity, thereby catalyzing chemical reactions. We applied our method to a Mn-doped $Co_3O_4$, as elaborated in the supplementary information. After training the XAS3Dabs model, we performed XANES fitting, a crucial step in the analysis process. We defined specific ranges for the atomic Cartesian coordinates, energy shift, and normalization factor as follows: [-0.3\AA,0.3\AA] for atomic Cartesian coordinates, [-10eV,10eV] for energy shift, and [0.7,1.3] for normalization factor. These ranges were chosen based on previous studies and our understanding of the system.
To determine the optimal fitting algorithm, we tested several optimization algorithms and compared their performance. We found that the Dividing Rectangles (DIRECT) algorithm\cite{DIRECT,DIRECT_L} delivered the best results, as shown in Fig.S10 in the supporting information. The DIRECT algorithm is a global optimization method that searches for the best solution by dividing the search space into smaller rectangles and exploring each one in detail. And it outperformed heuristic algorithms in terms of both accuracy and computational efficiency. The fitting went through 30000 steps.The fitting results showed that the average axial bond length of Mn is 0.3\AA longer than that of the plane bond length, indicating  the presence of a J-T effect in the local environment of doped Mn sites, whose trend is consistent with that described in the literature\cite{JT}. The fitting spectrum is shown in the Fig.\ref{fig:mnxanes}. It can be seen that all features of the experimental spectrum are reconstructed by fitting, including the shoulder peak at 6573eV and the scattering peak at 6616eV. Furthermore, we calculated the multiple scattering theoretical spectra based on the fitted 3D structure, and found that they matched well with the model predicted spectra, providing further support for our method. Single-spectrum calculations based on XAS3Dabs took approximately 0.2 seconds, including the normalization factor and energy shift fitting process, while a single multiple scattering calculation took 2.8 minutes (performed on a dual E5-2660v4 server with 28 parallel computing cores). This significant computational efficiency improvement compared to previous methods was achieved by leveraging advanced computational techniques and efficient custimized XAS3D model.
In summary, we applied our proposed method to the Mn-doped $Co_3O_4$ and verified its practicality in actual solid materials. This method holds significant potential for expanding the application of XANES fitting in high-degrees-of-freedom systems, particularly solid materials. The results of this study demonstrate the effectiveness of our method in fitting experimental XANES and extracting physical insights from complex materials.

\section{Conclusion and Outlook} 
This study presents a novel approach for XANES analysis, utilizing the XAS3Dabs graph neural network to compute XANES spectra and managing the fitting process through a global optimization algorithm. The effectiveness of this method is demonstrated in the $Fe_3O_4$, with further verification carried out in a Mn-doped $Co_3O_4$, which yields results that are consistent with existing literature data. One of the key advantages of this proposed method is that it does not require users to manually summarize structural parameters, thereby greatly simplifying the analysis process and expanding its potential application scope. This technique has significant implications for 3D structure analysis of solid materials, as well as for investigating structure-function relationships in fields such as energy and catalysis. Moreover, this method holds great promise for future development into an online 3D structure analysis tool for use with XAS-related beamlines. By streamlining the analysis process and increasing its accuracy and efficiency, this approach has the potential to revolutionize the field of XANES analysis and pave the way for new discoveries and advancements in material science.

\section{Funding information} 
We acknowledge financial support from the National Key Program of China (2020YFA0405800) and Platform of Advanced Photon Source Technology(PAPS), one of the Key Projects in the Planning of Huairou National Comprehensive Science Center, Beijing

\end{document}